\documentclass{acaces}

\newif\ifCOMMENTS
\COMMENTSfalse

\newcommand\yhc[2]{#2}

\usepackage{booktabs}
\usepackage{cleveref}
\usepackage{microtype}
\usepackage[sortcites=true,sorting=nyt,backend=biber]{biblatex}
\usepackage{tikz}
\usepackage[export]{adjustbox}

\setminted{fontsize=\small}

\newcommand\gsa{\emph{GSA}}
\newcommand\hgsa{\emph{h-GSA}}

\begin{document}

\title{GSA to HDL:\\
Towards principled generation of dynamically scheduled circuits}

\author{
Aditya Rajagopal\extranum{1},
Diederik Adriaan Vink\addressnum{1}\extranum{2},
Jianyi Cheng\addressnum{1}\extranum{2},
Yann Herklotz\addressnum{1}\extranum{2}
}

\address{1}{
  Imperial College London, UK
}

\extra{1}{E-mail: adityarajagopal0@outlook.com}
\extra{2}{E-mail: \{dav114,jc9016,ymh15\}@ic.ac.uk}

\pagestyle{empty}

\begin{abstract}
High-level synthesis (HLS) refers to the automatic translation of a software program written in a
high-level language into a hardware design.
Modern HLS tools have moved away from the traditional approach of static (compile time) scheduling
of operations to generating dynamic circuits that schedule operations at run time.
Such circuits trade-off area utilisation for increased dynamism and throughput.
However, existing lowering flows in dynamically scheduled HLS tools rely on conservative
assumptions on their input program due to both the intermediate representations (IR) utilised as
well as the lack of formal specifications on the translation into hardware.
These assumptions cause suboptimal hardware performance.
In this work, we lift these assumptions by proposing a new and efficient abstraction for hardware
mapping; namely \hgsa{}, an extension of the Gated Single Static Assignment (GSA) IR.
Using this abstraction, we propose a lowering flow that transforms \gsa{} into \hgsa{} and maps
\hgsa{} into dynamically scheduled hardware circuits.
We compare the schedules generated by our approach to those by the state-of-the-art
dynamic-scheduling HLS tool, Dynamatic, and illustrate the potential performance improvement from
hardware mapping using the proposed abstraction.
\end{abstract}


\vspace{-0.25em}
\section{Introduction and Related Works}
High Level Synthesis (HLS) tools convert a specification of a hardware circuit provided in a high
level programming language such as C/C++ into a hardware description language (HDL) that can then
be synthesised into the desired circuit.
Input programs in high level languages are often described with sequential semantics and
considerable control-flow as CPUs are highly optimised to execute such programs.
The challenge with HLS lies in successfully converting such a description into highly parallel
circuits.
To do so, most commercial tools such as Vivado HLS by AMD and Catapult HLS by Siemens utilise
static scheduling, i.e. the clock cycle at which each operation executes is determined at compile
time rather than runtime.
Static scheduling, which is sensitive to the latency of the operators, generates circuits
with predictable execution times, low resource consumption and high performance for input programs
with regular control-flow.
However, in programs with irregular control-flow, this approach can result in low throughput
\yhc{hardware designs}{} as \yhc{it}{static scheduling} relies on worst-case assumptions of
operation latencies in the absence of runtime information.

One solution to this issue is to generate circuits that execute purely based on dataflow, i.e. an
operation is only executed when all its inputs contain valid data from previous operations.
Such a circuit would execute in a manner that is insensitive to the latency of the components that
constitute it and would be able to execute each operation as early as possible.
\yhc{Towards}{Working towards} generating pure dataflow circuits, \cite{josipovic_dynamically_2018}
introduced Dynamatic, a dynamically scheduled HLS flow which bases its foundations \yhc{on work}{} on
latency insensitive circuit design \cite{carloni_theory_2001, carloni_methodology_1999}.
They demonstrate up to 2.5x improvement in execution time and 4x improvement in throughput on
programs with irregular control-flow when compared to statically scheduled HLS at the cost of up to
5x more resources.
Dynamatic uses a lowering flow that transforms a control-dataflow (CDFG) \footnote{dataflow within
basic blocks and control-flow between basic blocks} representation of the
program (in LLVM IR) directly to a dataflow \yhc{circuits}{circuit} by mapping IR constructs to predefined
elastic modules.
However, this process requires significant static analysis \cite{cheng_trets2023} to prevent
\yhc{the generation of artificial}{superfluous} control-flow dependencies that limit the amount of parallelism in the
generated hardware.

To avoid the pitfalls of mapping a CDFG to a dataflow circuit as well as the limitations of static
analysis from the lack of runtime information, \cite{unleashing_2022,
petersen_dynamically_nodate} propose transforming LLVM IR into new IRs that abstract dataflow into
the representation before lowering to a predefined set of elastic components.
\cite{unleashing_2022} propose to use Gated \yhc{Single Static}{Static Single} Assignment \cite{ottenstein_program_1990}
(GSA) while \cite{petersen_dynamically_nodate} propose the \emph{handshake} dialect in the MLIR
ecosystem.
However, \cite{unleashing_2022} neither provide formal semantics of the transformation nor the
resulting GSA, which makes it hard to reason about correctness of lowering. The
\emph{handshake} IR restricts the ability to perform further static analysis as it relies
explicitly on runtime information.
Instead, we propose a novel version of the GSA IR, \hgsa{}, that aims to alleviates both these
issues.
The proposed lowering flow first uses Compcert \cite{compcert_2009} to perform a formally verified
conversion of an SSA based IR to GSA \cite{herklotz_mechanised_2023}, then transforms this into
\hgsa{}, a version of GSA that is amenable to elastic circuit generation, and maps \hgsa{}
into a predefined set of elastic components.

The following sections demonstrate the increased dynamism that our methodology can generate over
Dynamatic and introduces the formal semantics of GSA nodes and invariants that hold when
transforming GSA to \hgsa{}.
Finally, we discuss the various future directions of research we envision this work will enable.

\vspace{-0.25em}
\section{Motivating Example}
\label{sec:motivating_eg}

\begin{figure}
\centering
\begin{subfigure}[b]{0.35\textwidth}
\begin{minted}[fontsize=\footnotesize]{c}
int foo() {
  int i, j, k;
  int a = 3, b = 5;
  L0: for (i=0; i<3; i++)
    a += f(i);
  L1: for (j=0; j<4; j++) {
    a += g(j);
    L2: for (k=0; k<3; k++)
      b += h(j, k); }
  return a + b; }
\end{minted}
\caption{Motivating example.}
\label{fig:motivating_source}
\end{subfigure}
\begin{subfigure}[b]{0.35\textwidth}
\centering
\includegraphics[height=10em]{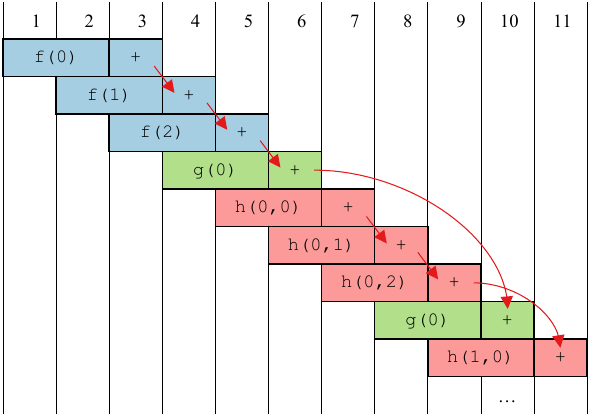}
\caption{Pipeline schedule by Dynamatic~\cite{josipovic_dynamically_2018}.}
\label{fig:base_schedule}
\end{subfigure}
\begin{subfigure}[b]{0.25\textwidth}
\centering
\includegraphics[height=10em]{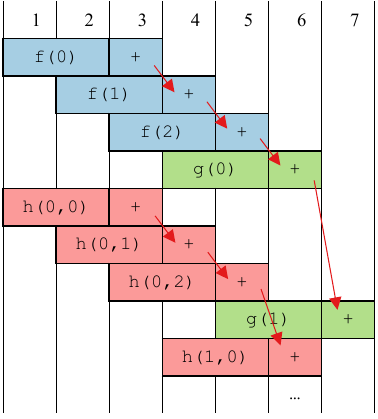}
\caption{Pipeline schedule by our work.}
\label{fig:opt_schedule}
\end{subfigure}
\caption{The red arrows mean the data dependence in the program. Existing dynamic scheduling restricts the loop iteration to {\em start} sequentially, although they can compute in parallel. Our work safely enables out-of-order loop computations by mapping GSA to hardware.}
\label{fig:motivation}
\end{figure}

This section motivates the use of GSA as in intermediate representation for HLS
by comparing the execution schedules generated by our tool vs. Dynamatic for a
simple motivating example (\Cref{fig:motivating_source}).
\Cref{fig:base_schedule} shows the schedule generated by Dynamatic.  As a
consequence of lowering directly from a CDFG (LLVM IR), Dynamatic generates pure
dataflow circuits within basic blocks, but requires basic blocks to start
sequentially. This means L2, which modifies variable \texttt{b} needs to start
after L1 (which increments \texttt{a} with \texttt{g(j)}) has started, even
though there is no data-dependency between \texttt{a} being incremented and
\texttt{b} being incremented. This results in reduced task level parallelism,
where the second iteration of L1, shown in green in the schedule, only starts on
cycle 8.

On the other hand, \Cref{fig:opt_schedule} shows the schedule that is achieved
by using \hgsa{}. This achieves the optimal schedule for this
example. As there are no basic-blocks in GSA, there are no
restrictions on when instructions start based on control-flow
dependencies. Instead, only data-dependencies are taken into account, leading to
all operations executing as soon as their dependencies are met. In this example,
lowering via h-GSA allows L0 and L2 to start at the same time. Due to the
write-after-write dependency between assign \texttt{a} in L0 and L1, L1 needs to
wait for L0 to finish before it can start, which is shown by the green loop
waiting for the blue loop to finish (\Cref{fig:opt_schedule}).  However,
contrary to the schedule provided by the Dynamatic circuit, L2 (red) can start
executing at the first cycle, because there is no data-dependency between L0 and
L2.  This additional dynamism is a feature of GSA because it provides a purely
dataflow centric view of the whole program using the primitive functions that
are further described in \cref{sec:formalisms}.


\vspace{-0.25em}
\section{IR Level Formal Semantics and Hardware Mapping}
\label{sec:formalisms}

This section will describe the notable differences between \hgsa{} and the
version of GSA~\cite{ottenstein_program_1990} formalised in
CompCert~\cite{herklotz_mechanised_2023}.  Starting from such a formal
description means that there is a well-understood starting point on which we can
base the translation from GSA into dynamically scheduled circuits.  GSA is an
extension to SSA, defining two types of gating $\phi$-instructions, namely
$\mu$-instructions at loop headers and $\gamma$-instructions at the exit of
conditional statements.  Additional nodes called $\eta$-instructions are also
added at loop exits.  These nodes can then be assigned different semantics based
on their usual control-flow (software) semantics, and in \hgsa{}, dataflow
(hardware) semantics, which are shown in \Cref{tab:semantics-gsa}.

Our work performs a lowering from this GSA representation to a netlist of
pre-defined components.  Each operation in the software program is translated
into a hardware component and connected using the handshake interface
\cite{josipovic_dynamically_2018}.  The arithmetic and logic operations used by
our work are the same as the ones in Dynamatic, however, we add additional
hardware modules which implement the hardware semantics of the GSA-specific IR
constructs.

\begin{table}[tb]
  \centering
  \caption{Semantics of GSA functions in both hardware and software. These
    functions do not perform computation and only forward inputs to outputs when
  specific conditions are met.}\label{tab:semantics-gsa}
  \begin{tabular}{lp{6.5cm}p{6.5cm}}\toprule
    \textbf{Instruction}
    & \textbf{Software Semantics}
    & \textbf{Hardware Semantics} \\ \midrule

    $d \leftarrow \mu(r_0, r_{\rm i})$
    & Behave like a $\phi$-instruction from SSA and inspect the incoming
      control-flow edge to choose the register to assign to $d$.
    & Stateful component accepting inputs from $r_0$ the first time, then
      accepting inputs from $r_{\rm i}$ until it is reset.\\

    $d \leftarrow \gamma(\overrightarrow{(p_i, r_i)})$
    & Make a local choice of which register $r_i$ to pick based on the
      evaluation of its predicate $p_i$.
    & Component that selects the $r_i$ based on the evaluation of $p_i$. \\

    $d \leftarrow \eta(p, r)$
    & Only allow execution of the semantics to proceed when $p$ evaluates to true.
    & Component stalls until $p$ is true. It also resets its
      corresponding $\mu$ instruction to accept new inputs. \\ \bottomrule

  \end{tabular}
  \label{tab:gsa-semantics}
\end{table}

Interesting modifications had to be performed to the GSA program produced by
CompCertGSA before we could generate functional hardware from it.  Even though
the transformation to GSA is formally verified, the hardware requires stronger
properties about the structure of the GSA to execute correctly.  Firstly,
additional guarantees are needed about registers not appearing after they have
been gated.  Secondly, each $\mu$ instruction has to be paired with at least an
$\eta$ instruction so that it is reset upon loop exit. Finally, particular patterns
need additional GSA nodes (described below), where the arrows represent a
control-flow graph, $\mu$-$\eta$ pairs define entry and exit of loops, and P and C are producers and
consumers of values. 

\vspace{0.5em}
\begin{tabular}[htb]{cp{12.5cm}}
  \adjustbox{valign=t}{\begin{tikzpicture}[every node/.style={font=\footnotesize}]
    \node[draw,circle] (p) {C};
    \node[draw,circle,above left of=p,xshift=-0.5em] (c) {P};
    \node[above of=p] (n1) {$\mu$};
    \node[below of=p] (n2) {$\cdot$};
    \draw[->] (p) -- (n2);
    \draw[->] (n1) -- (p);
    \draw[->] (n2) to [out=40,in=320] (n1);
    \draw[<-] (n1) -- (c);
  \end{tikzpicture}}
  & When a producer is outside of the loop that contains its consumer, then
$\mu$ nodes needs to be generated for it.  This is not the case in software
    because the value assigned by the producer will always be accessible by
    reading its register, but in a dynamic circuit it needs to be passed around
    the loop ($\mu$ generates the same value at each new loop iteration).\\[0.5em]
  \adjustbox{valign=t}{\begin{tikzpicture}[every node/.style={font=\footnotesize}]
      \node[draw,circle] (p) {P};
      \node[above right of=p] (loophead) {$\mu$};
      \node[below right of=loophead] (eta) {$\eta$};
      \node[draw,circle,above right of=eta] (c) {C};
      \draw[->] (p) -- (loophead);
      \draw[->] (loophead) -- (eta);
      \draw[->] (eta) -- (c);
      \draw[->] (loophead) to [out=130,in=50,loop] (loophead);
  \end{tikzpicture}}
  & One also needs to guarantee that any consumers that are separated from their
  producer by a loop have an appropriate $\mu$ and $\eta$ pair.  This rule in
    particular is quite hardware specific, because through control-flow
    execution of the program this pair would not be required, as the value of
    the producer P remains unchanged, and can be read by the consumer.  In
    dynamic hardware it is needed to continually feed values through the
    circuit. \\
\end{tabular}




\vspace{-0.25em} 

\section{Conclusion and Future works}
\label{sec:future_works}

In conclusion, using GSA as an IR for a dynamically scheduled HLS tool
demonstrates promise towards improving upon Dynamatic.  In addition to that,
building upon precisely specified semantics reduces the gap towards a verified
translation from software to hardware, making it possible to build
correct-by-construction dynamic circuits. It is currently difficult to verify
the correctness of a dynamic circuit specifically because it is
latency-insensitive, leading to a notion of correctness which specifies that
``eventually'' the outputs should be consistent.  There are many topics that
still need to be addressed however, such as efficient handling of memory and automatic
buffering of circuits.  These are features that Dynamatic already supports, and
would be needed to carry out a fair performance comparison.

\printbibliography
\end{document}